\documentclass[11pt]{elsarticle}

\usepackage{xcolor}
\usepackage{soul}

\usepackage{geometry}
\journal{Journal of Theoretical Biology}

\usepackage{graphicx}
\usepackage{amsmath,amssymb}
\usepackage{subfigure}

\newcommand{\bi}{\begin{itemize}}
\newcommand{\ei}{\end{itemize}}
\newcommand{\be}{\begin{enumerate}}
\newcommand{\ee}{\end{enumerate}}
\newcommand\ba{\begin{eqnarray}}
\newcommand\ea{\end{eqnarray}}
\newcommand\bc{\begin{center}}
\newcommand\ec{\end{center}}

\newcommand\ra{\rightarrow}
\newcommand{\nn}{\nonumber}

\newcommand{\bfx}{\ensuremath{\mathbf{x}}}
\newcommand{\bfy}{\ensuremath{\mathbf{y}}}
\newcommand{\bfA}{\ensuremath{\mathbf{A}}}
\newcommand{\bfB}{\ensuremath{\mathbf{B}}}

\newcommand{\bbR}{\ensuremath{\mathbb{R}}}
\newcommand{\e}{\ensuremath{\mathrm{e}}}
\newcommand{\xA}{x_a}
\newcommand{\xB}{x_b}
\newcommand{\xC}{x_c}
\newcommand{\I}{\uppercase\expandafter{\romannumeral1}}
\newcommand{\II}{\uppercase\expandafter{\romannumeral2}}
\newcommand{\III}{\uppercase\expandafter{\romannumeral3}}
\usepackage{color}

\begin{document}

\begin{frontmatter}

\title{Demographic noise slows down cycles of dominance}


%
%
\author[mymainaddress]{Qian Yang}
\ead{Q.Yang2@bath.ac.uk}

\author[mymainaddress]{Tim Rogers\corref{mycorrespondingauthor}}
\cortext[mycorrespondingauthor]{Corresponding author}
\ead{T.C.Rogers@bath.ac.uk}

\author[mymainaddress]{Jonathan H.P. Dawes}
\ead{J.H.P.Dawes@bath.ac.uk}

\address[mymainaddress]{Centre for Networks and Collective Behaviour, 
							Department of Mathematical Sciences, \\
							University of Bath,
							Bath BA2 7AY, UK}

\begin{abstract}
We study the phenomenon of cyclic dominance in the paradigmatic Rock--Paper--Scissors model, as occurring in both stochastic individual-based models of finite populations and in the deterministic replicator equations. The mean-field replicator equations are valid in the limit of large populations and, in the
presence of mutation and unbalanced payoffs, they exhibit an attracting limit cycle. The period of this cycle depends on the rate of mutation; specifically, the period grows logarithmically as the mutation rate tends to zero. We find that this behaviour is not reproduced in stochastic simulations with a fixed finite population size. Instead, demographic noise present in the individual-based model dramatically slows down the progress of the limit cycle, with the typical period growing as the reciprocal of the mutation rate. Here we develop a theory that explains these scaling regimes and delineates them in terms of population size and mutation rate. We identify a further intermediate regime in which we construct a stochastic differential equation model describing  the transition between stochastically-dominated and mean-field behaviour. 
\end{abstract}


\end{frontmatter}


\section{Introduction}
\label{intro}

Many mathematical models in ecology are well-known to be capable of generating oscillatory dynamics in time; important examples stretch right back to the initial
work of Lotka and Volterra on predator-prey interactions
\cite{may_nonlinear_1975,guckenheimer1988,krupa1997,britton_2005,nowak_evolutionary_2006,toupo_nonlinear_2015}. 
Such models, although dramatic
simplifications when compared to real biological systems, have
a significant impact in shaping our understanding of the
modes of response of ecological systems and are helpful in
understanding implications of different strategies for, for
example, biodiversity management, and the structure of
food webs \cite{reichenbach_coexistence_2006}. 

Competition between species is a key driver of complex dynamics in ecological models. Even very simple competitive interactions can yield complex
dynamical behaviour, for example the well documented
example of the different strategies adopted by 
three distinct kinds of side-blotched lizard \cite{sinervo_rockpaperscissors_1996}.
Similar cyclical interactions occur in bacterial
colonies of competing strains of {\it E. coli}
\cite{kerr_local_2002,kirkup_antibiotic-mediated_2004,weber_chemical_2014}. In mathematical neuroscience dynamical
switches of this type have been
referred to as `winnerless competition' since there is no best-performing
state overall \cite{rabin2001,tsai2013}. 

Evolutionary Game Theory (EGT) provides a useful framework for modelling competitive interaction, in particular the replicator equations \cite{taylor1978evolutionary,schuster1983replicator} give a dynamical systems interpretation for models posed in game-theoretic language. Work by many authors, including in particular
Hofbauer and Sigmund
\cite{hofbauer1988,hofbauer1994}
has resulted in a very good understanding of replicator
equation models for competing species. Recent
work has extended these deterministic approaches to consider
stochastic effects that emerge from consideration of
finite, rather than infinite, populations. The classic Rock--Paper--Scissors (RPS) provides an important example of stochastic phenomena in ecological dynamics. When mutation (allowing individuals to spontaneous swap strategies) is added to the replicator equations  for the RPS game, the deterministic can exhibit damped oscillations that converge to a fixed point. In \cite{mobilia_oscillatory_2010}, it was shown that stochastic effects present in finite populations cause an amplification of these transient oscillations, leading to so-called quasi-cycles \cite{mckane2005predator}. For smaller values of mutation rate, the deterministic system passes through a Hopf bifurcation, and a limit cycles appears. Some past studies exist on the role of noise around limit cycles, such as \cite{boland2008limit,boland2009limit}, in which small-scale fluctuations around the mean-field equations are explored using Floquet theory. More recently, it has been discovered that noise can induce much stronger effects including counterrotation and bistability \cite{newby_effects_2014}. 

In this paper we combine deterministic and stochastic approaches in order to present a comprehensive description of the effect of demographic fluctuations around cycles of dominance in the RPS model. We determine three regimes, depending on the scaling of population size $N$ and mutation rate $\mu$. The basic link between stochastic individual-based dynamics and population-level ODEs is a theorem of Kurtz \cite{kurtz_strong_1978}, allowing us to construct a consistent set of individual-level behaviours corresponding to the mean-field replicator dynamics for the RPS model that we take as our starting point. Between these two views of the dynamics lies a third: the construction
of a stochastic differential equation (SDE) that captures the transition
between them. Changing variables to the asymptotic phase of the ODE limit cycle reveals that the contribution of the stochasticity is to speed up some parts of the phase space dynamics and to slow down
others but that the overall effect is to markedly increase
the oscillation period. Our central conclusion is that
as the stochastic effects become more important, the period
of the oscillations increases rapidly, and this slowing down
is a significant departure from the prediction of oscillation
periods made on the basis of the mean-field ODE model.

The structure of the remainder of the paper is as follows.
In section~\ref{sec:models} we introduce the replicator
dynamical model for the rock-paper-scissors game with mutation.
The mean-field ODE version of the model is well-known and we derive
a self-consistent individual-based description; this is not as
straightforward as one might initially imagine. We show numerically
that the two models give the same mean period for the cyclic dynamics when
the mutation rate is large but disagree when $\mu$ is small.

In section~\ref{sec:region1} we summarise the computation to
estimate the period of the limit cycle when $\mu$ is small. This follows
the usual approach, dividing up trajectories into local behaviour near
equilibrium points, and global maps valid near the unstable manifolds 
of these saddle points.
Section~\ref{sec:region3} turns to the stochastic population model and
analyses the dynamics in terms of a Markov chain. This leads to 
a detailed understanding of the individual-level behaviour in the limit of
small mutation rate $\mu$.
Section~\ref{sec:region2} then
fills the gap between the analyses of sections~\ref{sec:region1}
and~\ref{sec:region3} by deriving an SDE that allows us to understand
the relative contributions of the stochastic behaviour and the deterministic
parts in an intermediate regime.
Finally, section~\ref{sec:discussion} discusses our results and concludes.

\section{Models for Rock-Paper-Scissors with mutation}
\label{sec:models}

\subsection{Rock-paper-scissors with mutation}
\label{model:RPS}


Rock--Paper--Scissors (RPS) is a simple two-player, three-state game which
illustrates the idea of cyclic dominance: a collection of strategies,
or unchanging system states, in which each state in turn is unstable to
the next in the cycle. In detail: playing the strategy `rock' beats the strategy `scissors' but loses to the strategy `paper'; similarly, `scissors' beats `paper' but loses to `rock'. When the two players play the same strategy the contest is a draw.


This information is summarised in the payoff matrix
\begin{equation}
\label{nonzero_payoff}
P:=
\left(
\begin{array}{ccc}
0        &  -1-\beta       & 1 \\
1 &  0       &  -1-\beta \\
-1-\beta        & 1 & 0 
    \end{array}
		\right)
   \end{equation}
where $\beta \geq 0$ is a parameter that indicates that
the loss incurred in losing contests is greater than the payoff gained
from winning them. When $\beta=0$, the row and column sums of
$P$ are zero: this is the simplest case. 
When $\beta>0$, the game becomes more complicated, particularly
when we would like to relate the behaviour at the population level
to the individual level, as we discuss later in
sections~\ref{model:odes} and~\ref{model:chemical}.

\subsection{Deterministic rate equations}
\label{model:odes}

Setting the RPS game in the context of Evolutionary Game Theory (EGT),
one considers a large well-mixed population of $N$ players playing the game
against opponents drawn uniformly at random from the whole
population.
We are then interested in the
proportions of the total population who are playing different
strategies at future times. The state of the system
is given by the population fractions
$(\xA(t),\xB(t),\xC(t)):=(N_A(t),N_B(t),N_C(t))/N$ where
$N_{A,B,C}(t)$ are the numbers of players playing strategies
A, B and C respectively.

The proportions
of the population playing each strategy are expected to change over time
according to the typical payoff received, as compared to the average over the whole population. The simplest mean field model for the resultant dynamics are the replicator equations
\begin{equation}
\label{eqn:ODE_1}
\dot{x_i}=x_i(t)\left(\sum_{j}P_{ij}x_j(t)-\sum_{j,k}P_{jk}x_{j}(t)x_{k}(t)\right),
\end{equation}
where the subscripts $i,j$ take values in $\{a,b,c\}$ and the proportions
$x_i$ sum to unity. 
    
A common variant of the model introduces the additional mechanism of
mutation between the three strategies, occurring between any pair with equal frequency. Mutation 
affects the rate of change of strategy $i$ over time since
the strategies other than $i$ will contribute new players of $i$
at rates $\mu$ while $i$ will lose players at a rate given by $2\mu x_i$
as these players switch to a different strategy.

In the particular case of Rock--Paper--Scissors, the combined effects of the replicator dynamics together with
mutations between strategies gives rise to the ordinary differential equations
\begin{equation}
\label{eqn:rate_eq}
\begin{split}
\dot{x}_a & =  \xA [\xC -(1+\beta)\xB + \beta(\xA \xB + \xB \xC + \xA \xC)]+\mu(\xB + \xC -2\xA),\\
\dot{x}_b & =  \xB [\xA -(1+\beta)\xC + \beta(\xA \xB + \xB \xC + \xA \xC)]+\mu(\xA + \xC -2\xB),\\
\dot{x}_c & =  \xC [\xB -(1+\beta)\xA + \beta(\xA \xB + \xB \xC + \xA \xC)]+\mu(\xA + \xB -2\xC).
\end{split}
\end{equation}
We note that these equations are to be solved in the region of
$\mathbb{R}^3$ where all coordinates are non-negative. This region is
clearly invariant under the vector field~\eqref{eqn:rate_eq}. Moreover,
the constraint $\xA + \xB + \xC=1$ is required to hold at all times.
    

The system~\eqref{eqn:rate_eq} possess a single interior equilibrium point $x^{*}=(1/3,1/3,1/3)$, in which the three strategies are balanced. Straightforward linear stability analysis shows that this equilibrium is stable when $\mu>\mu_c=\beta/18$. Previous work by Mobilia \cite{mobilia_oscillatory_2010} has shown that
the system undergoes a Hopf bifurcation 
as $\mu$ is lowered and that for $\mu<\mu_c$
trajectories of~\eqref{eqn:rate_eq}
spiral away from $x^*$ and are attracted to
a unique periodic orbit which is stable, i.e a limit cycle.


\subsection{Stochastic chemical reactions}
\label{model:chemical}

The replicator equations shown above are expected to hold in the limit of infinitely large population size. 
In finite populations, however, the behaviour of many competing individuals is
more appropriately modelled as a Markov process describing the random timing individual events. It is common 
practice to specify such a stochastic model as a chemical reaction scheme, which, chosen appropriately should recover the ODEs~\eqref{eqn:rate_eq} in the limit of large systems. It is interesting to note that different
stochastic individual-based models may give rise to the same mean-field ODEs,
so that the question of constructing a stochastic reaction scheme starting from
a particular set of ODEs does not have a unique answer. Moreover, the construction
of the stochastic model is subject to a number of natural constraints, for example that all reaction rates are at all times non-negative.




With the goal of studying dynamics around the fixed point $x^*$ a reaction scheme was proposed in~\cite{mobilia_oscillatory_2010} based on consideration of the frequency-dependent Moran model with rates chosen to match the mean-field equations as required in the infinite system limit. Although that scheme is well motivated and perfectly correct for the regimes previously studied, it will not be suitable for our study of the stochastic dynamics around the limit cycle, as the reaction rates can take negative (unphysical) values near the system boundaries. Thus our first task here is to choose a reaction scheme that permits us to study the whole of state space in the case $\beta>0$, $\mu\ll1$. 

For the ``replicator" part of the dynamics we propose the chemical reactions
\begin{align}
     \label{reaction_dc}
    A+B\xrightarrow{1} B+B,\quad
    B+C&\xrightarrow{1}C+C,\quad
    C+A\xrightarrow{1}A+A, \\
    \label{reaction_beta_ab}
     A+B+B&\xrightarrow{\beta} B+B+B ,\\
    \label{reaction_beta_ac}
     A+A+C&\xrightarrow{\beta}A+A+A, \\
    \label{reaction_beta_bc}
    B+C+C&\xrightarrow{\beta}C+C+C,
    \end{align}
Mutations are included in the model through the additional six reactions
     \begin{equation}
        \label{rule_mutation_2}
        \begin{split}
            A\xrightarrow{\mu}B, &\quad B\xrightarrow{\mu}C, \quad C\xrightarrow{\mu}A, \\
            A\xrightarrow{\mu}C, &\quad B\xrightarrow{\mu}A, \quad C\xrightarrow{\mu}B.
        \end{split}
    \end{equation}
This information can be more usefully summarised in terms of an integer `jump matrix' $\mathbf{S}$ describing the changes to $\mathbf{x}$ caused by the various reactions (so element $S_{ij}$ represents the increment or decrement of species $i$ taking place in reaction $j$), and vector $\mathbf{r}(\mathbf{x})$ describing the reaction rates. For the reactions in~\eqref{reaction_dc} we have
\begin{equation}
    \mathbf{S_1}=\left(\begin{array}{cccccccccccc}
    -1 & 0 & 1  & -1  & 1  & 0 \\
    1 & -1 & 0  & 1  & 0  & -1 \\
    0 & 1 & -1  & 0  & -1  & 1 \\
    \end{array}\right),
     \end{equation}
     \begin{equation}
     \begin{split}
 \mathbf{r}_1=(&x_ax_b,x_bx_c,x_cx_a,\beta x_ax_b^2, \beta x_a^2x_c, \beta x_bx_c^2)^{T}.
     \end{split}
     \end{equation}
Whilst the mutation reactions in~\eqref{rule_mutation_2} are specified by
\begin{equation}
    \mathbf{S_2}=\left(\begin{array}{cccccccccccc}
    -1 & 0 & 1  & -1  & 1  & 0 \\
    1 & -1 & 0  & 0  & -1  & 1 \\
    0 & 1 & -1  & 1  & 0  & -1 \\
    \end{array}\right),
     \end{equation}
     \begin{equation}
     \begin{split}
 \mathbf{r}_2=(&x_a,x_b,x_c,x_a,x_b,x_c)^{T}.
     \end{split}
     \end{equation}
The jump matrix and rate vector for the full scheme are found simply by concatenation:
\begin{equation}
\mathbf{S}=(\mathbf{S_1},\mathbf{S_2})\,,\quad \mathbf{r}=\left(\begin{array}{c}\mathbf{r}_1\\\mathbf{r}_2\end{array}\right)\,.
\label{Sr}
\end{equation}

We choose to express our reactions in this form in order to appeal to a very useful theorem of Kurtz \cite{kurtz_strong_1978}, stating that in the limit $N\to\infty$ of large populations, the stochastic process described by these reactions converges to the deterministic dynamical system $$\dot{\mathbf{x}}=\mathbf{S} \mathbf{r}(\mathbf{x})\,.$$ It is easy to check that the above scheme thus reproduces the system \eqref{eqn:rate_eq}. Although we will develop our analysis for this particular chemical reaction scheme, it is important to reiterate that it is not unique in reproducing the replicator equations in the large population limit. However, we argue that the important features of the stochastic slowdown we observe will be common to all reasonable models, including those with more biologically realistic properties such as involving only two-body interactions, and permitting varying population sizes.

\subsection{Simulations}
\label{model:simu}

To gain an initial insight into the differences between the deterministic and
stochastic viewpoints for the RPS game with mutation,we use the Gillespie algorithm \cite{gillespie1977exact},
the standard and widely used stochastic simulation algorithm (SSA),
to simulate the chemical reactions \eqref{reaction_dc}-\eqref{rule_mutation_2} in order
to illustrate the typical dynamics in the stochastic case and to compare
that with the deterministic case.

\begin{figure}[!ht]
\bc
\includegraphics[width=0.45\textwidth]{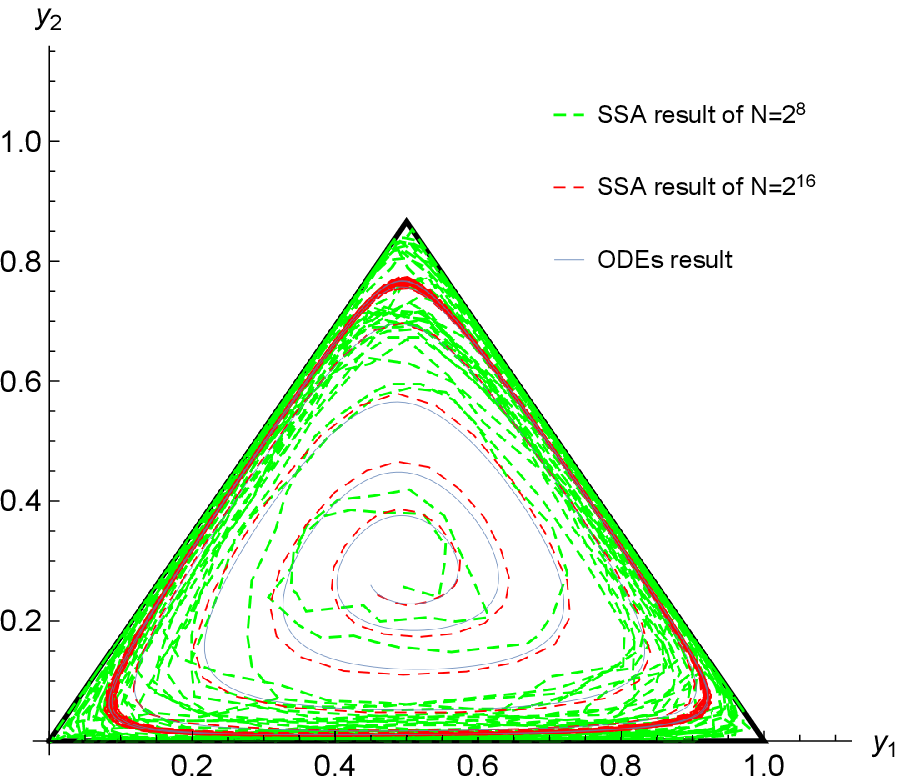}\hspace{1cm}\includegraphics[width=0.45\textwidth]{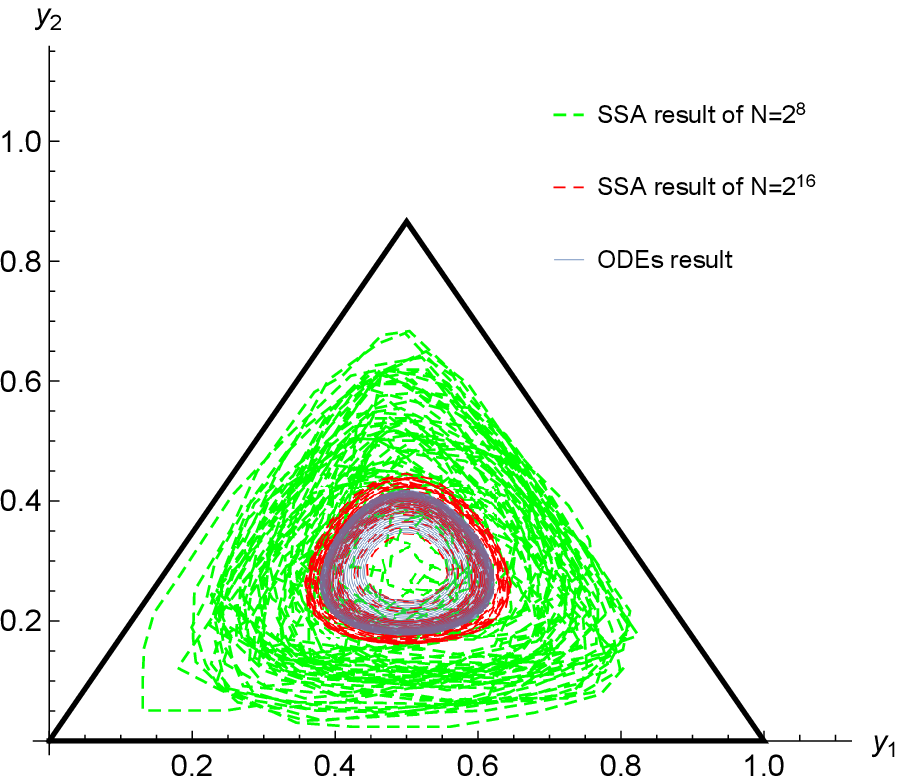}
(a)\hspace{6cm}(b)

\caption{Illustrative comparisons between trajectories of the ODEs (blue solid
lines) and realisations of the stochastic simulations (red and green dashed lines).
`Unbalanced' means that the payoff matrix is not zero-sum, i.e. $\beta>0$.
(a) When $\mu \ll \mu_c$ the trajectories lie close to the boundary of
the phase space, $\beta=1/2$, $\mu=1/216$. (b) When $\mu$ is only
slightly smaller than $\mu_c$, trajectories lie much closer to the central
equilibrium point, $\beta=1/2$, $\mu=5/198$.}
\label{fig:formation_of_stable_cycle}
\ec
\end{figure}

Figure~\ref{fig:formation_of_stable_cycle} presents results comparing
the stochastic and deterministic cases for two different values of 
the mutation rate $\mu$. In each plot we show a typical realisation of
the stochastic simulation algorithm, for two different finite, but large,
population sizes $N=2^8$ (green dashed line) and $N=2^{16}$ (red dashed line),
together with a trajectory of the ODEs (blue solid line).

The solution to the ODEs is shown by the blue line in~\ref{fig:formation_of_stable_cycle}(a).
The red dashed line in~\ref{fig:formation_of_stable_cycle}(a) starts from
a very similar initial condition and evolves 
similarly: spiralling out towards the boundary of phase space and at large times
occupying a region of phase space near to the limit cycle 
but with small fluctuations around it.
The green dashed line in~\ref{fig:formation_of_stable_cycle}(a) shows much
larger fluctuations around the limit cycle, including excursions that take
the simulation onto the boundaries of the phase space, and much closer to
the corners. Note that because of the
mutations, the boundaries of the phase space are not absorbing states for
the random process (or invariant lines for the ODE dynamics).

Figure~\ref{fig:formation_of_stable_cycle}(b) illustrates the behaviour for
a significantly larger value of $\mu$ for which the limit cycle for the ODEs
(blue curve) lies much closer to
the centre of the phase space. The SSA for $N=2^{16}$ lies close to the limit
cycle but fluctuates around it; for $N=2^8$ the fluctuations are much larger
and the sample path of the stochastic process lies outside the limit cycle
for a large proportion of the simulation time.

To quantify the differences between the three results shown
in each part of figure~~\ref{fig:formation_of_stable_cycle}, we
focus on one specific aspect of the dynamics: the period of the oscillations
around the central equilibrium point at $(1/3,1/3,1/3)$.
For the ODEs, the period of the limit cycle can be defined to be the
smallest elapsed time between successive crossings of a hyperplane
in the same direction, for example the plane
$x_a=1/2$ (this is a sensible choice because, as we will later show, the effect of noise is smallest at this part of the cycle). We denote by $T_{\text{ODE}}$ the period of the deterministic
limit cycle. 
In stochastic simulations, a trajectory may by chance cross a hyperplane back and forth several times in quick succession, thus the time between crossings may not represent a full transit of the cycle. We avoid this complication by measuring the period as three times the transit time between a hyperplane and its $2\pi/3$ rotation. The expected value of this oscillation period in the stochastic case we denote by
$T_{\text{SSA}}$.
    
\begin{figure}[!ht]
\bc
\includegraphics[width=0.9\textwidth]{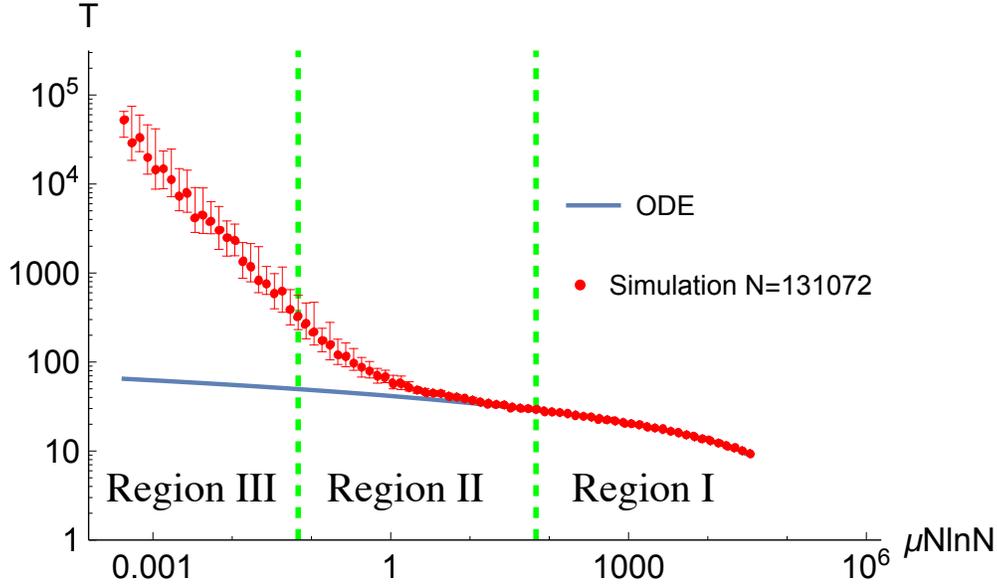}
\caption{Comparison between the period $T_\text{ODE}$ of the limit cycle in the
deterministic case and the average period $T_\text{SSA}$ of oscillations
in the stochastic case, as a function of the parameter $\mu$
at fixed $N=2^{17}$. The blue line shows $T_{\text{ODE}}$ and red dots indicate 
$T_{\text{SSA}}$. As $\mu$ decreases, the two periods start to separate
from each other in Region II where $\mu N\log N \approx 1$.}
\label{fig:threeregions}
\ec
\end{figure}

Figure~\ref{fig:threeregions} provides a quantitative comparison of the
different dependencies of the average period $T_{\text{SSA}}$
of the stochastic simulations and 
the period $T_{\text{ODE}}$ of the deterministic ODEs on the mutation rate $\mu$, for
$0<\mu<\mu_c$. 
The blue solid line indicates the relatively slow increase in ther period
$T_{\text{ODE}}$ as $\mu$ decreases.
The red error bars indicate the range of values of the oscillation
period in the stochastic case, with the averages of those values shown by the
red dots. The data in this figure for the stochastic simulations was obtained 
from simulations at a fixed value $N=2^{17}$ with $\mu$ varying, but
for presentational reasons that will become clearer later in the paper we have
chosen to plot the scaled quantity $\mu N \ln N$ on the horizontal axis.
We then observe that if $\mu$ is sufficiently small then
the difference between $T_{\text{ODE}}$ and $T_{\text{SSA}}$ is very significant:
the oscillations in the stochastic simulation have a much
longer period, on average, than that predicted by the ODEs, while if $\mu N \ln N$ is 
larger than unity, the agreement in terms of the oscillation period, between the
deterministic and stochastic simulations is very good. We label these
two regimes `Region \III' and `Region \I' respectively. The cross-over
region where $\mu N \ln N \approx 1$ we label `Region \II'.

In the following sections of the paper we will focus on each of the
three regions in turn. In section \ref{sec:region1} we study
the period of the limit cycle in the ODEs
(region \I). In section \ref{sec:region3} we
analyse the stochastic dynamics to determine the average period of cycles in
region \III. In section \ref{sec:region2} the
cross-over between the deterministic and stochastic regions is
understood through the analysis of an SDE that combines
both deterministic and stochastic effects, thus describing
region \II.

\section{Analysis of the periodic orbit in region I}
\label{sec:region1}

\begin{figure}[!ht]
\bc
\includegraphics[width=0.8\textwidth]{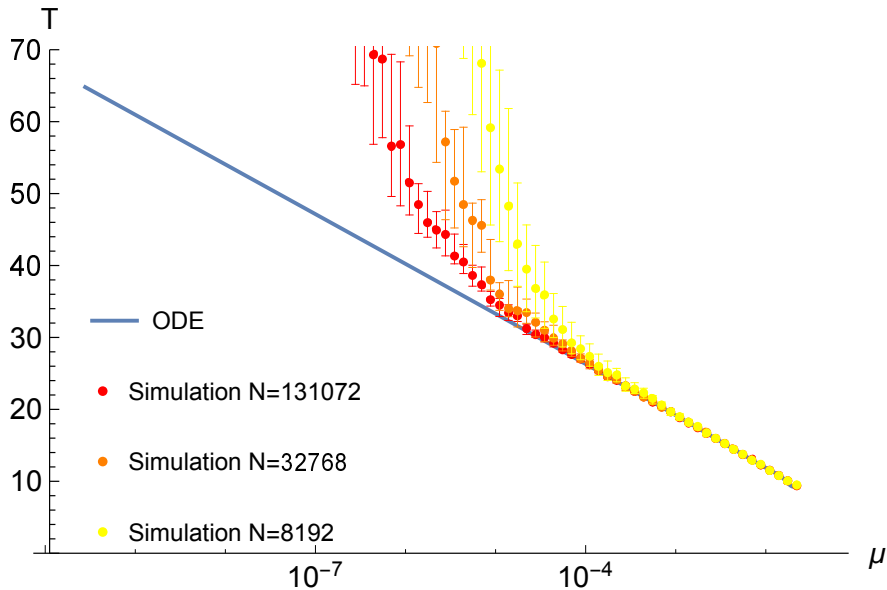}
\caption{Period $T_\text{SSA}$ for three different values of $N$, compared
with the result for $T_\text{ODE}$ (blue solid line) on a log-linear
plot indicating the scaling law $T_{\text{ODE}}\propto -3\log \mu$ which applies
in region \I\ where the
stochastic simulations behave in a similar way to the ODEs.}
\label{TODELogLinearPlot}
\ec
\end{figure}

Region \I\ is defined
to be the right hand side of figure~\ref{fig:threeregions}; more
precisely, it is the regime in which $N \gg 1$ and
$(N\log N)^{-1} \ll \mu <\mu_c$.
In this region, stochastic simulation results show only very small
fluctuations about a mean value, and this mean value coincides well with
numerical solutions to the ODEs.
This is evidence that the behaviour of the deterministic ODEs provides a
very good guide to the stochastic simulation results in this region.
Hence our goal in this section is to explain the asymptotic result that the
period $T_{\text{ODE}}\propto -3\log \mu$ for $\mu$ small, but for large,
even infinite, $N$ as shown in figure~\ref{TODELogLinearPlot}.

\begin{figure}[h!]
\bc
\includegraphics[width=0.8\textwidth]{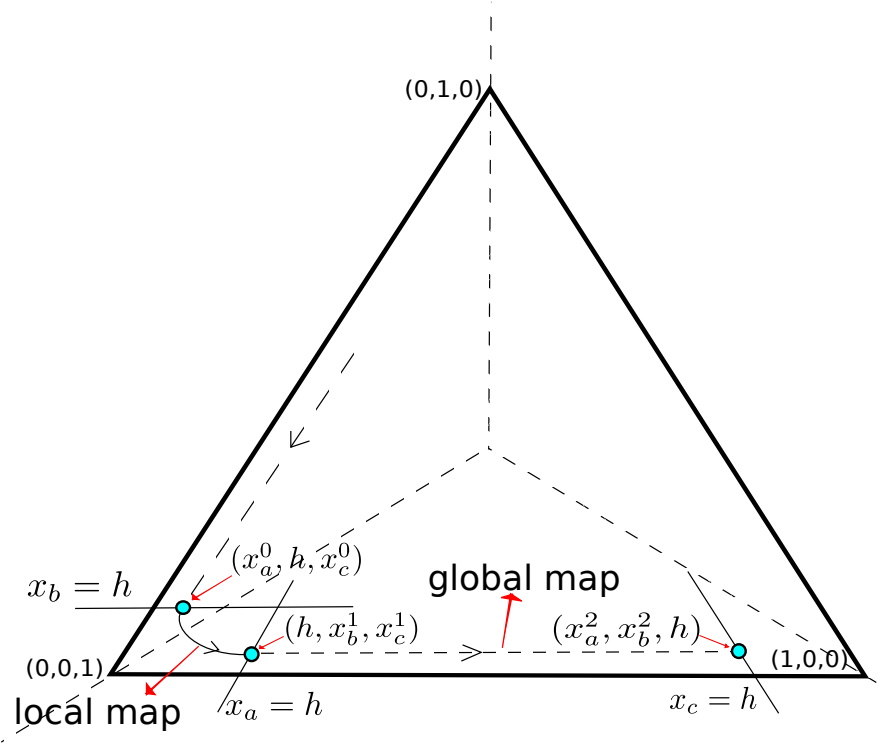}
\caption{This picture shows the local map and the global map starting from initial position $(x_a^{0},h,x_c^{0})$.}
\label{fig:localmap_globalmap}
\ec
\end{figure}

The standard approach to the analysis of trajectories near the limit cycle,
in the regime where it lies close to the corners
of the phase space, is to construct local maps that analyse the flow near the
corners, and global maps that approximate the behaviour of trajectories close
to the boundaries \cite{guckenheimerholmes}.  
Figure \ref{fig:localmap_globalmap} shows the construction of a local map
in the neighbourhood of the corner where $x_c=1$, following by the global
map from this neighbourhood to a neighbourhood of the corner where $x_a=1$. 
In this section we consider local and global maps in turn, in sections~\ref{sec:local}
and~\ref{sec:global}, and then in
section~\ref{sec:compose} we study their composition, and
deduce an estimate for the period $T_\text{ODE}$ of the limit cycle.

\subsection{Local map}
\label{sec:local}

Using the notation as in figure~\ref{fig:localmap_globalmap}, we begin by
defining a neighbourhood of the corner $x_c=1$ by setting $0<h \ll 1$ to be 
a small positive constant.
We assume that the trajectory for the ODEs~\eqref{eqn:rate_eq} starts at
the point $(x_a^{(0)},h,x_c^{(0)})$ at time $t=0$, and 
arrives at the point $(h,x_b^{(1)},x_c^{(1)})$ at time $t=T_1>0$. 
For the whole time $0<t<T_1$ the trajectory lies close to the corner
$(0,0,1)$, so we suppose that $0<x_a,x_b,u\ll 1$ where $u:=1-x_c$. Note that
we have $x_a+x_b-u=0$ since $x_a+x_b+x_c=1$.

Then the behaviour of the ODEs~\eqref{eqn:rate_eq} in this neighbourhood is very similar
to that of their linearisation obtained by dropping terms higher than linear order
in the small quantitites $0<x_a,x_b,u\ll 1$. For the linearised system
we obtain
\begin{equation}
\label{eqn:rate_eq2}
\left\{
\begin{aligned}
\dot{x}_a & =\mu-x_a(3\mu-1),\\
\dot{x}_b & =\mu-x_b(1+\beta+3\mu),\\
\dot{u}   & =-x_b(1+\beta)+x_a+2\mu-3\mu u.
\end{aligned}
\right.
\end{equation}
The equations \eqref{eqn:rate_eq2} are linear and constant coefficient and so
can be solved analytically; note that the first and second equation are actually
decoupled from each other and from the $u$ equation.
Integrating from $t=0$ to the time $t=T_1$, we denote
the point on the trajectory by $(x_a(T_1),x_b(T_1),x_c(T_1))\equiv
(x_a^{(1)}= h, x_b^{(1)}, x_c^{(1)})$ where
\begin{align}
x_b^{(1)} & = \frac{\mu}{\gamma}\left(1 - \e^{-\gamma T_1}\right) + h \e^{-\gamma T_1}, 
\label{eqn:rate_eq3} \\
x_a^{(1)} & = h =\frac{\mu}{\hat{\gamma}} + \e^{-\hat{\gamma}T_1}\left(x_a^{0}
-\frac{\mu}{\hat{\gamma}}\right) \label{eqn:rate_eq4}
\end{align}
where $\gamma:=1+\beta+3\mu$ and $\hat{\gamma}:=3\mu-1$.
Equation~\eqref{eqn:rate_eq4} allows us to express $T_1$ in terms of $x_a^{(0)}$:
\begin{equation}
\label{T_1}
T_1=-\frac{1}{\hat{\gamma}}\log\left(\frac{h-\mu/\hat{\gamma}}{x_a^{(0)}-\mu/\hat{\gamma}}\right),
\end{equation}
and we can now use~\eqref{T_1} to eliminate $T_1$ from~\eqref{eqn:rate_eq3} and
obtain a relationship between $x_a^{(0)}$ and $x_b^{(1)}$ which takes the form
\begin{equation}
\label{x_a_1_new}
x_b^{(1)}=\frac{\mu}{\gamma}+\left(h-\frac{\mu}{\gamma}\right)
\left(\frac{h-\mu/\hat{\gamma}}{x_a^{(0)}-\mu/\hat{\gamma}}\right)^{\gamma/\hat{\gamma}}.
  \end{equation}
This relationship is the key part of the local map near the point $(x_a,x_b,x_c)=(0,0,1)$
that we will use in what follows.

\subsection{Global map}
\label{sec:global}

For the global map, we observe that trajectories remain 
close to one of the boundaries (in this case, the boundary $x_b=0$) and so we
propose that the trajectory starting from 
$(x_a^{(1)}= h, x_b^{(1)}, x_c^{(1)})$
arrives at the point $(x_a^{(2)},x_b^{(2)},h)$ at time $t=T_2$. Referring
to~\eqref{eqn:rate_eq}, the ODE for $x_b$ near the boundary can be well approximated
by taking just $\dot{x}_b=\mu$ (since $x_a+x_c=1$ when $x_b=0$), so its solution is
\begin{equation}
x_b^{(2)}=x_b^{(1)}+(T_2-T_1)\mu =x_b^{(1)}+\mathcal{C}_0\mu,
\label{x_a_2}
\end{equation}
where we denote the elapsed time by $\mathcal{C}_0:=T_2-T1$.
As is typical in these analyses, trajectories take a relatively short time
to arrive at the hyperplane $x_c^{(2)}=h$ starting from $x_c^{(1)}$,
compared to the time taken to move along the part of the trajectory
from $x_c^{(0)}$ to $x_c^{(1)}$; 
this is intuitively because the absolute value of $\dot{x}_c$ on the global
part of the map between $T_1$ and $T_2$ is much larger than on the local
part, i.e. when $0 < t < T_1$. 
As a result, the time taken on the global part of the map,
$\mathcal{C}_0:=T_2-T_1$ is much less than the local travel time $T_1$;
the majority of the time spent on the limit cycle is taken up with travel near the
corners.

\label{sec:compose}

The composition of local and global maps near the corner $x_c=1$ and boundary $x_b=0$
is now straightforward: we combine~\eqref{x_a_1_new} and~\eqref{x_a_2} to obtain:
\begin{equation}
x_b^{(2)}=\mathcal{C}_0\mu+
\frac{\mu}{\gamma}+\left(h-\frac{\mu}{\gamma}\right)\left(\frac{h
-\mu/\hat{\gamma}}{x_a^{(0)}-\mu/\hat{\gamma}}\right)^{\gamma/\hat{\gamma}}.
\end{equation}

We can now use the permutation symmetry inherent in the dynamics to complete
the analysis. Due to the fact that the model is rotationally symmetric, the next stage of the evolution is local map again with the same parameter values. The trajectory
will start from $(x_a^{(2)},x_b^{(2)},h)$ and stay near the corner $(1,0,0)$ for long time
before arriving at a point, say, $(x_a^{(3)},h,x_c^{(3)})$ where we will construct another
global map, and so on:
\ba
\label{map_1}
x_a^{(0)}\xrightarrow{local} x_b^{(1)}\xrightarrow{global} x_b^{(2)}
\xrightarrow{local} x_c^{(3)} \xrightarrow{global} x_c^{(4)}
\xrightarrow{local} x_a^{(5)} \xrightarrow{global} x_a^{(6)}
\rightarrow \cdots
\ea
which can be summarised further as
\ba
\label{map_2}
x_0\xrightarrow{\text{\quad}local\text{ }\&\text{ }global\text{\quad}}
x_1\xrightarrow{\text{\quad}local\text{ }\&\text{ }global\text{\quad}}
x_2\xrightarrow{\text{\quad}local\text{ }\&\text{ }global\text{\quad}}
x_3 \rightarrow \cdots
\ea
Equations~\eqref{map_1} and~\eqref{map_2} above define a one-to-one correspondence
between the points $\{(x_a^{(n)},x_b^{(n)},x_c^{(n)})\}$ on a trajectory and a
sequence of values (selecting appropriate coordinates)
$\{x_n\}$. From the previous discussion on local and global maps, 
the map that generates the sequence $\{x_n\}$ takes the form
\begin{equation}
\label{sequence_1}
x_{n+1}=\mathcal{C}_0\mu+
\frac{\mu}{\gamma}+\left(h-\frac{\mu}{\gamma}\right)\left(\frac{h-\mu/\hat{\gamma}}{x_n-\mu/\hat{\gamma}}\right)^{\gamma/\hat{\gamma}},
\end{equation}
where, as before, $\gamma:=1+\beta+3\mu$ and $\hat{\gamma}:=3\mu-1$.

If iterates of the map \eqref{sequence_1} converge to a fixed point then
this corresponds to a stable limit cycle for the ODE dynamics.
We now estimate the location of
this fixed point and deduce an estimate for the period of the resulting limit cycle.

Let $y_n:=x_n/\mu$ be a scaled version of $x_n$, then~\eqref{sequence_1} can be written as:
\ba
y_{n+1} & = & \mathcal{C}_0+
\frac{1}{\gamma}+\left(h-\frac{\mu}{\gamma}\right)\frac{1}{\mu}\left(\frac{\mu y_n+\frac{\mu}{1-3\mu}}{h+\frac{\mu}{1-3\mu}}\right)^{\frac{1+3\mu+\beta}{1-3\mu}} \nn \\
      & = & \mathcal{C}_0+
\frac{1}{\gamma}+\left(h-\frac{\mu}{\gamma}\right)\mu^{\frac{1+3\mu+\beta}{1-3\mu}-1}\left(\frac{ y_n(1-3\mu)+1}{h(1-3\mu)+\mu}\right)^{\frac{1+3\mu+\beta}{1-3\mu}},
\label{sequence_2}
\ea
where $\gamma:=1+3\mu+\beta$.
Denoting the fixed point of the map by $y^{*}$, from \eqref{sequence_2} we see
that $y^{*}= \mathcal{C}_0 + 1/\gamma+o(\mu)$ in the limit $\mu \to 0$.
This observation is crucial in order to ensure that we obtain the correct leading-order
behaviour and distinguish carefully between the various small quantities in the problem.
Then it follows that $x^{*}=\mu y^{*}\approx \mathcal{C}_1\mu$, where
$\mathcal{C}_1=\mathcal{C}_0 + 1/({1+\beta})$, in the limit $\mu\to 0$.

Introducing this leading-order approximation for $x^{*}$ into~\eqref{T_1}, we obtain
an estimate of the time spent in a neighbourhood of the corner $T_1$ of the stable limit cycle:
\begin{equation}
\label{T_1_final}
T_1=\frac{1}{1-3\mu}\log\left({\frac{h+\mu(1-3\mu)}{\mathcal{C}_1\mu+\mu(1-3\mu)}}\right).
\end{equation}
Since in this case $\mu$ is assumed to be very small, \eqref{T_1_final}
takes the form, at leading order,
\begin{equation}
\label{T_1_final_2}
T_1=\frac{1}{1-3\mu}\log\left({\frac{1}{\mu}}\right) +\mathcal{B}_0 \text{ as } \mu\to0.
\end{equation}
where $\mathcal{B}_0$ is a constant.

Finally, as remarked on above, because trajectories remain near each corner for
large parts of the period of the orbit, the local map travel time $T_1$ is the
dominant contribution, compared to the time spent on the global map.
Hence the period of the limit cycle is given at leading order by considering only
the contribution from the three local maps required in one full period of the
limit cycle. Hence our estimate for $T_\text{ODE}$ becomes
\begin{equation}
\label{period_limit_cycle}
T_\text{ODE} \approx 3T_1 = -3\log\mu+\mathcal{B}_1, 
\end{equation}
as $\mu \to 0$.

\section{Analysis of the periodic orbit in region III}
\label{sec:region3}

\begin{figure}[!ht]
\bc
\includegraphics[width=0.8\textwidth]{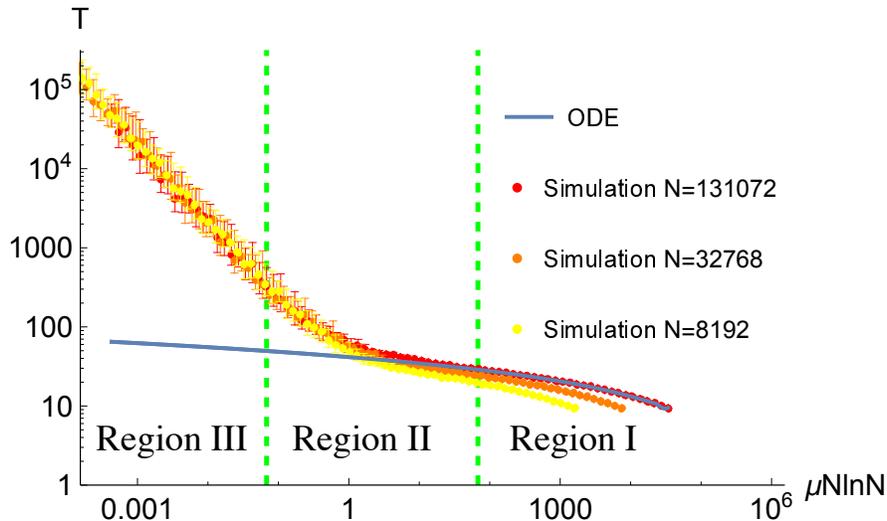}
\caption{Log-log plot of the period $T_\text{SSA}$ as a function of the mutation rate $\mu$
for three different values of $N$, showing that in
region \III\ the mean period scales roughly
as $T_\text{SSA} \approx (\mu N \ln N)^{-1}$.}
\label{fig:TwithDifferentN}
\ec
\end{figure}

We now turn our attention to Region III, where the period of the orbit in the
stochastic simulations increases
much more rapidly as $\mu$ decreases, at fixed finite $N$, than the prediction from
the analysis of the ODEs in section~\ref{sec:region1} above suggests.
Figure~\ref{fig:TwithDifferentN} illustrates this by plotting the period $T_\text{SSA}$
as a function of $\mu N \ln N$ for three different values of $N$. By plotting, on
 a log-log scale,
the mean values of the periods and the error bars from an ensemble of stochastic
simulations we observe that the period $T_\text{SSA} \approx (\mu N \ln N)^{-1}$
for small $\mu$,
with a constant that does not demonstrate any systematic dependence on $N$. In fact,
the range of values of $N$ presented here is small: we cannot distinguish from
these numerical results the precise form of the dependence on $N$.

As well as the dependence of the period $T_\text{SSA}$ on $\mu$ it is also of
interest to determine the extent of Region III in which this scaling behaviour
applies; in other words how small is $\mu$ required to be in order to
move into this regime? The numerical data in figure~\ref{fig:TwithDifferentN} indicate
that the cross-over from the ODE result to this new stochastic scaling
arises when $\mu N\log N = 1$.

In this section, then, our aim is to explain firstly why this new scaling regime in Region III exists, and secondly, why it extends as far as $\mu N \ln N =1$. We will find
that in fact the period should scale as $T_\text{SSA} \approx (\mu N)^{-1}$
for small $\mu$, but that the observation that the cross-over occurs when
$\mu N \ln N =1$ is correct; this explains why we have chosen to plot
figure~\ref{fig:TwithDifferentN} in the form that it appears.

\begin{figure}[h!]
\bc
\includegraphics[width=0.5\textwidth]{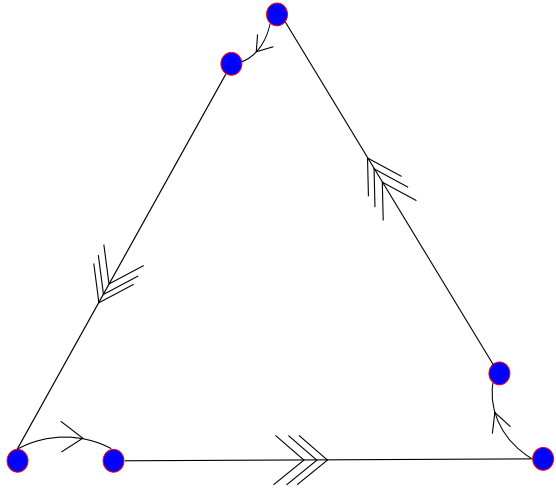}
\caption{Sketch of the dynamics in the small-$\mu$ stochastic limit.
The system is strongly attracted to corner states where the population
is all in one state. Mutation is then the only mechanism for escape, and the
expected time for the first mutation (curved arrows) is longer than the
expected time required for subsequent steps along a boundary towards the next
corner (straight lines with arrows).}
\label{Hitting_time_cal}
\ec
\end{figure}

Careful examination of the numerical simulations in this regime show that their
behaviour is qualitatively different from that in Region I discussed above.
In stochastic simulations the system becomes strongly attracted to the corner
states (which would be absorbing states in the absence of mutation) and then
can only escape from a corner when a mutation occurs; mutations are rare when
$\mu$ is small. Our analysis later in this section shows that the mean period of oscillation
is dominated by the contribution from the time needed to escape one step from
a corner. Trajectories then typically move along a boundary towards the next corner.
We show below that, although this latter part requires at least $N-1$ steps,
the expected total time required is less than the waiting time to escape from the corner.
Our schematic approach is illustrated in
figure~\ref{Hitting_time_cal} while sketches this separation into a first
step away from a corner, following by movement along the adjoining boundary.
    
In the following subsections we consider in detail three issues: firstly, in
section~\ref{sec:prob} we show that the probability that the motion
along a boundary is towards the next corner in the sequence, and that
the systems state hits this next corner with a probability
that tends to $1$ as $N \rightarrow \infty$. Since we are
interested in the regime where the mutation rate is very small,
we carry out this calculation in the limit $\mu=0$. In section~\ref{sec:time}
we compute the expected time until the system hits this next corner, again
setting $\mu=0$. Finally,
in section~\ref{sec:mutation} we compare this expected time for the system
state to evolve along the boundary with the expected time until a mutation
occurs. Together, this analysis confirms the intuitive picture outlined in
the previous paragraph.


\subsection{Probability of hitting the absorbing status}
\label{sec:prob}

In this section we consider the state-discrete and time-continuous Markov Chain dynamics of the system evolving
along a boundary, ignoring the effect of mutation. In this case the
system is a one dimensional chain of states labelled 0 to $N$, corresponding to, for example, the number of strategy $B$ individuals invading a population of size $N$ initially composed of all strategy $A$ individuals. Let $h_i=\mathbb{P}_i(\text{hit }0)$ be
the probability that the system hits (and is therefore
absorbed by) state $0$ having started at node $i$, and similarly let
$\bar{h}_i=\mathbb{P}_i(\text{hit }N)$ be the probability of hitting state
$N$ having started at node $i$.
%

When the system is in state $i$ there are two possible moves: jumps to the left
or to the right.
The rate at which jumps to the left occur is
$p_i={(N-i)i^2}/{N^2}$; the rate at which jumps to the right occur is
$q_i={(N-i)i^2(1+\beta)}/{N^2}+{(N-i)i}/{N}$. The transition
probabilities of moving to the left and to the right are then
$\ell_i={p_i}/{(p_i+q_i)}$ and $r_i={q_i}/{(p_i+q_i)}$; clearly $\ell_i+r_i=1$.
We remark that the ratio of transition probabilities can be simplified
to be
\ba
\label{left_right_ratio}
\frac{\ell_i}{r_i}=\frac{p_i}{q_i}=\frac{1}{(1+\beta)+\frac{N}{i}}. 
\ea
This allows us to derive a recurrence relation for the probability $h_i$
that the system hits the state $0$ starting from state $i$:
\begin{equation}
\label{rf_1}
\left\{
\begin{aligned}
& h_0=1,\\
& h_i=\ell_i h_{i-1}+r_i h_{i+1},\text{\quad for } i=1\dots N-1, \\
& h_N=0.\\
\end{aligned}
\right.
\end{equation}
Recurrence relations of this type are straightforward to solve by standard methods \cite{norris_markov_1998}. In our case we find 
\begin{equation}
h_1 = \frac{\sum_{i=1}^{N-1}\gamma_i}{1+\sum_{i=1}^{N-1}\gamma_i},
\end{equation}
and of course $\bar{h}_1=1-h_1$. Here we define $\gamma_i$ as 
\begin{equation}
\gamma_i: =\prod_{j=1}^{i}\frac{\ell_j}{r_j} = 
\prod_{j=1}^{i}\frac{j}{j(1+\beta)+N}<\frac{i!}{N^i} 
\end{equation}
We now examine the limiting behaviour of this result and
prove that $h_1\to 1$ as $N\to \infty$.
Since
\begin{equation}
0 < \gamma_{i+1} = \frac{p_{i+1}}{q_{i+1}}\gamma_i, \nn
\end{equation}
and ${p_{i+1}}/{q_{i+1}}<1$, it follows that
$\gamma_{i+1} < \gamma_{i}$. Also, since $\gamma_i <{i!}/{N^i}$, we
conclude that
\ba
0 < \sum_{i=1}^{N-1}\gamma_i & = & \gamma_1+\sum_{i=2}^{N-1}\gamma_i  
\nn \\ & \leq & 
\frac{1}{N}+(N-2)\frac{2}{N^2}, \nn
\ea
which clearly tends to zero as $N\ra \infty$. Hence
\begin{equation}
\label{lim_to_0}
\lim_{N\to \infty} \sum_{i=1}^{N-1}\gamma_i = 0.
\end{equation}
In conclusion, we have shown that, given that it starts at state $1$,
the probability of the system hitting the absorbing state at $N$ tends to $1$
as $N \ra \infty$.

\subsection{Average hitting time}
\label{sec:time}

Having shown that the system reaches state $N$ with high probability, we now
consider the expected time until this happens.
Let $T_i$ be is the first time at which the system hits an absorbing
state starting from state $i$ at time $0$, i.e. the time at which the system
hits either $0$ or $N$. Define $\tau_i=\mathbb{E}(T_i)$ for $0 \leq i \leq N$,
the expected hitting time starting from state $i$.
Clearly we have that $\tau_0=\mathbb{E}(T_0)=0$ and $\tau_N=\mathbb{E}(T_N)=0$.
Through a similar calculation to setion~\ref{sec:prob}, we now
compute $\tau_1$, the mean time taken to reach either $0$ or $N$, starting
from $1$.

The recurrence formula for expected hitting times $\tau_i$ can be
computed as the sum of the expected time spent in state $i$ before jumping
either left or right, plus the expected future time required when in that
new state:
\begin{equation}
\label{rf_2}
\tau_i=\frac{1}{p_i + q_i} + \ell_i \tau_{i-1} + r_i \tau_{i+1}, \text{ \quad for }i=1\dots N-1.
\end{equation}
This three-term recurrence can be rearranged to give a two term recurrence on the variable $\tau_{i-1}-\tau_i$, whose solution yields
\begin{equation}
\label{result_hitting_time_4}
\tau_1 = {\left(1+\sum_{i=1}^{N-1}\gamma_i\right)}^{-1}
\sum_{i=1}^{N-1}\left(\frac{1}{q_i}\sum_{j=i}^{N-1}\frac{\gamma_j}{\gamma_i}\right).
\end{equation}
We now wish to examine the asymptotic behaviour of this expression
for the expected hitting time, in the limit when $N \gg 1$.
Initial numerical explorations lead us to propose that $\tau_1 \propto \ln{N}$
when $N$ is sufficiently large. In the remainder of this section we will
deduce this estimate systematically.

First, note that from~\eqref{lim_to_0} we know that
\begin{equation}
\label{lim_to_1}
\lim_{N\to \infty} \left(1+\sum_{i=1}^{N-1}\gamma_i\right) = 1.
\end{equation}

From the discussion before \eqref{left_right_ratio} we have that
\ba
q_i & = & i(N-i)\frac{1}{N}\left[\frac{i}{N}(1+\beta) + 1 \right]  \nn \\
\Rightarrow \frac{1}{q_i} & = & \frac{N}{i(N-i)}\frac{1}{1+(1+\beta)\frac{i}{N}}.
\label{eqn:qi}
\ea
Since
\ba
\frac{1}{2+\beta} < \frac{1}{1+(1+\beta)\frac{i}{N}} < 1, \nn
\ea
(as $1 \leq i \leq N-1$), we can bound $\sum_{i=1}^{N-1}{1}/{q_i} $ as follows
\begin{equation}
\label{eqn:result1}
\begin{split}
\frac{1}{2+\beta} \sum_{i=1}^{N-1}\frac{N-i+i}{(N-i)i} \leq
& \sum_{i=1}^{N-1}\frac{1}{q_i} 
\leq \sum_{i=1}^{N-1}\frac{N-i+i}{(N-i)i}  \\
\implies \frac{2}{2+\beta} \ln (N-1) \leq
& \sum_{i=1}^{N-1}\frac{1}{q_i}  \leq 2 \left( 1+ \ln (N-1) \right), 
\end{split}
\end{equation}

Next we prove that, for any $1 \leq i \leq N-1$,
$\sum_{j=i}^{N-1}\gamma_j/\gamma_i$ is bounded as $N \to \infty$.
From the definition of $\gamma_j$ we see that
\begin{equation}
\frac{\gamma_j}{\gamma_i}=\prod_{k=i+1}^{j}\frac{1}{(1+\beta)+\frac{N}{k}}<\prod_{k=i+1}^{j}\frac{k}{N+k}=\frac{j!}{i!}\frac{(N+i)!}{(N+j)!}.
\end{equation}
Therefore,
\begin{equation}
\sum_{j=i}^{N-1}\frac{\gamma_j}{\gamma_i} < 
\sum_{j=i}^{N-1}\frac{j!}{i!}\frac{(N+i)!}{(N+j)!}
= \frac{1}{N-1}\left[i+N-2N\frac{N!(N+i)!}{i!(2N)!}\right]\leq2. \nn
\end{equation}
We note that if we set $i=N-1$ in the above, then the limiting value is small, due
to the influence of the negative term that is a ratio of factorials, but the limit
must still be positive;
in the case $i=1$ we have a limit that is closer to $1$. In all cases,
since $1 \leq i \leq N-1$ the limiting value must be non-negative, and at most $2$.
Hence we see that
\ba
\lim_{N \to \infty} \sum_{j=i}^{N-1}\frac{\gamma_j}{\gamma_i} & \leq & 2. \label{eqn:result2}
\ea
We now apply the results~\eqref{lim_to_1}, ~\eqref{eqn:result1} and~\eqref{eqn:result2} to~\eqref{result_hitting_time_4} in order to deduce the result
\begin{equation}
\tau_1 \leq 4 \left( 1+ \ln (N-1) \right)
\end{equation}
which is an upper bound on the expected time, starting in state $1$, until the system hits one of the two absorbing states $0$ or $N$.
Together with the conclusion of the previous section, in which we computed that
the probability that the system arrives in state $N$ tends to $1$ as $N \to \infty$,
we can conclude that the expected time required for the system to hit state $N$
is no greater than $4 (1 + \ln N)$. In using this result later, we
will omit the subdominant constant term $1$ since we are concerned primarily with
values of $N$ for which $\ln N \gg 1$, in the limit $N \to \infty$.

\subsection{The effect of mutation on the stochastic dynamics}
\label{sec:mutation}

The analysis in the previous two subsections ignored the role of mutation in order
 to understand the dynamics on the boundary of phase space and, in particular, to
estimate the time required to travel along the boundary to a corner.

Since in the absence of mutation the corners are absorbing states, mutation
plays an important role in moving the system from a corner (state $0$) to
state $1$ on the boundary, allowing it then to travel further towards the next corner.
The mutation rate $\mu$, together with our assumptions on the stochastic dynamics,
imply that, for any system state,
the time until the next mutation event $M$ occurs is exponentially
distributed:
\begin{equation}
\mathbb{P}(M>t)=e^{-\mu Nt} \nn
\end{equation}
For system states on the boundary of phase space, we would expect
that mutations would move them away, into the interior, where the analysis in
the previous sections might become less useful. This is unlikely to
happen if mutations are not expected during the time taken for the system
to move along the whole boundary, i.e. if
\begin{equation}
\label{eq_mu_hit_1}
\mathbb{P}(\text{hit state }N\text{ before mutation})=\mathbb{P}(M>4\ln N)=\e^{-4\mu N \ln N}
\end{equation}
is close to $1$.
From the form of~\eqref{eq_mu_hit_1}, if $N$ is fixed and $\mu\to 0$, then
this probability of hitting state $N$ before any mutation occurs tends to 1, which
implies that the system remains in boundary states and the analysis of
sections~\ref{sec:prob} and~\ref{sec:time} applies. On the other hand, if
$N\to \infty$, and $\mu$ remains fixed, then mutation is expected to occur
before the system reaches the next corner, and so the system state
tends to leave the boundary. The intermediate balance between these
two regimes occurs when $\mu \sim (N \ln N)^{-1}$.

An equivalent discussion can be framed in terms of the sketch
of the dynamics indicated in figure~\ref{Hitting_time_cal}.
The time required for a full period of the oscillation is composed of
two contribution on each boundary piece:
the first contribution $\tau_\text{m}$ is the time required to jump, via mutation, from a 
corner to a state with one new individual of the appropriate kind. This mutation
takes an expected time $\tau_\text{m} \sim 1/(N\mu)$ since there
are $N$ individuals and each mutates independently at a rate $\mu$.
The second contribution $\tau_1$ is the time required to traverse the boundary starting
from state $1$.
This is approximately $\tau_1 =4\ln N$. If $\tau_\text{m}\gg \tau_1$, i.e.
$\mu \ll (4N\ln N)^{-1}$, then the largest contribution to the
mean period of oscillation is from the mutation events, and so we expect
in this regime to have the period of the orbit being dominated by
the time required for three independent mutations to occur, i.e.
$T_{SSA} \sim 3/(N\mu)$.

\section{Analysis of the periodic orbit in region II}
\label{sec:region2}
 
Region \II\ is the cross-over
region between `large $\mu$' where the ODE approximation is valid, and
`small $\mu$' where the stochastic approach based on a Markov Chain, is
appropriate.
Using the analysis of the previous section we see mathematically speaking that
region \II\ arises where $\mu\sim(N\ln N)^{-1}$. 
In region \II\ the stochastic simulations show large fluctuations around the
ODE predictions, but the system does not spend time always near the boundaries
of phase space, so it is not clear that the analysis of region \III\
should apply directly. 

In this section we examine region \II, and explain why the fluctuations in the
stochastic system act to increase the period of the oscillations rather than
to decrease it. This involves a third approach to the dynamics, using a stochastic
differential equation derived from the 
chemical reaction model and which is valid for large, but not infinite, system sizes. 
We show that the SDE
approach captures, in some detail, the transition between the deterministic
and the fully stochastic regimes described in previous sections of the paper.

\subsection{Stochastic differential equation}
\label{sec:sde}

It was proved by Kurtz \cite{kurtz_strong_1978} that trajectories of Markov jump processes specified in terms of a stoichiometric matrix $\mathbf{S}$ and rate vector $\mathbf{r}(\mathbf{x})$, as discussed in section~\ref{model:chemical}, are well approximated for large $N$ by the trajectories of the SDE
\begin{equation}
\label{eqn:sde}
\mathrm{d}\mathbf{x}=\mathbf{A}(\mathbf{x})\mathrm{d}t
+ \frac{1}{\sqrt{N}} \mathbf{G}(\mathbf{x})\mathrm{d}\mathbf{W}(t)\,.\\
\end{equation}
Here $\mathrm{d}\mathbf{W}_t$ is a vector, each
element of which is an independent Wiener process \cite{gardiner_stochastic_2009}, and the vector $\mathbf{A}$ and matrix $\mathbf{G}$ are given by 
\begin{equation}
\mathbf{A}(\mathbf{x})=\mathbf{S}\mathbf{r}(\mathbf{x})\,,\quad \mathbf{G}_{ij}(\mathbf{x})=\mathbf{S}_{ij}\sqrt{\mathbf{r}_j(\mathbf{x})} \,.
\end{equation}
In the calculation of the statistical properties of this equation, it is more useful to consider the matrix
$\mathbf{B}=\mathbf{G}\mathbf{G}^{T}$. For our system we obtain the explicit formulas
\begin{equation}
\label{FPE:particle_drift}
\begin{split}
\mathbf{A}_1 & = x_a[x_c-(1+\beta) x_b 
+\beta(x_a x_b + x_b x_c + x_a x_c)] + \mu(x_b + x_c - 2x_a), \\
\mathbf{A}_2 & = x_b[x_a-(1+\beta) x_c 
+\beta(x_a x_b + x_b x_c + x_a x_c)] + \mu(x_a + x_c - 2x_b),        
\end{split}
\end{equation}
\begin{equation}
\label{FPE:particle_diffusion}
\begin{split}
\mathbf{B}_{11} & = x_a x_b +x_c x_a + (x_a x_b^2 + x_c x_a^2)(2+\beta) +\mu(2x_a+x_b+x_c),\\
\mathbf{B}_{12} & = \mathbf{B}_{21}  =-(x_a x_b +x_a x_b^2(2+\beta)+\mu(x_a+x_b)),\\
\mathbf{B}_{22} & =x_a x_b + x_b x_c + (x_a x_b^2 + x_b x_c^2)(2+\beta) +\mu(x_a+2x_b+x_c).
\end{split}
\end{equation}

\subsection{Asymptotic phase of points near the periodic orbit}
\label{sec:phidot}

In order to explain the increase in the period of the orbit as
the fluctuations grow, we use the SDE~\eqref{eqn:sde} to derive
an equation for the angular velocity around the limit cycle, and then
compute the period of the limit cycle by integrating
the angular velocity. In this section we extend this idea by deriving
an SDE for the angular velocity. This allows us to investigate the effects
of noise on the period of the orbit. A fully analytic approach
is unfortunately not possible, so our approach is a combination of
numerical and analytic methods. Identifying the correct scalings for
features of the limit cycle however enables us to confirm the
various asymptotic scalings found in regions \I\ and \III\ and to see
how they both contribute in this region, region \II.

\begin{figure}[!ht]
\bc
\includegraphics[width=0.9\textwidth]{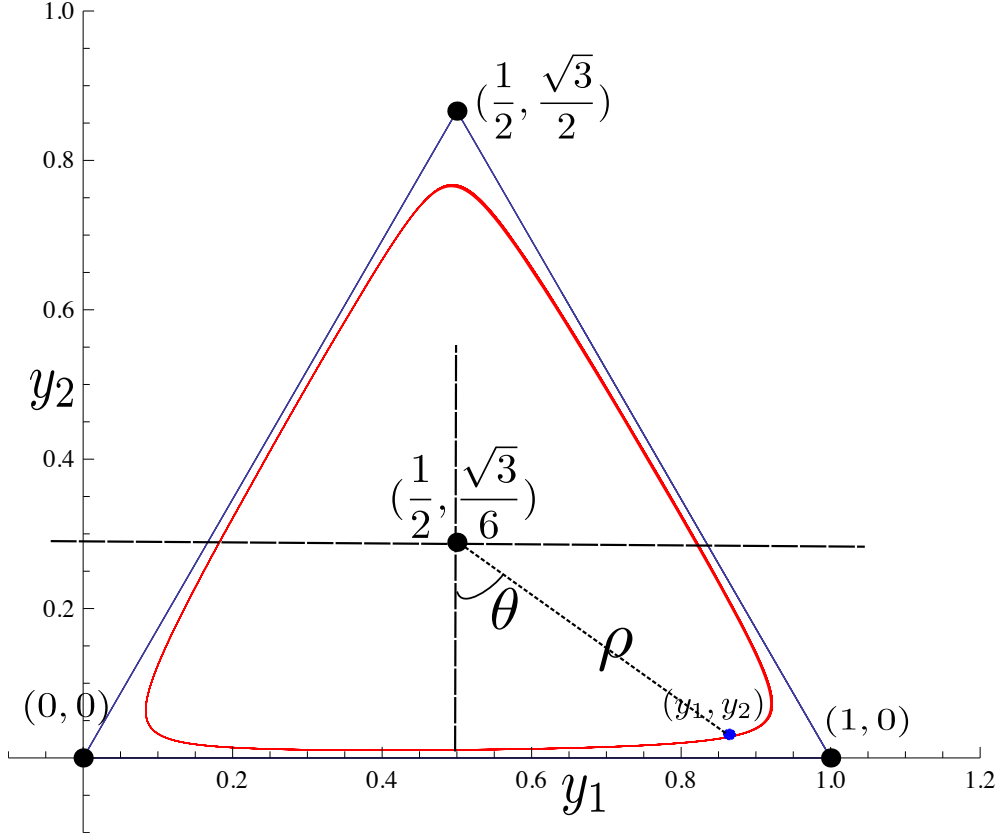}
\caption{The red circle is the limit cycle when $\beta=1/2, \mu={1}/{216}$, the blue triangle is the limit cycle when $\beta=1/2,\mu=10^{-6}$. When $\mu$ is very small, it is hard to tell the difference between triangle boundary and limit cycle without zooming-in.}
\label{limit_cycle_example}
\ec
\end{figure}

Since the phase space is two dimensional, we follow the presentation started
in section \ref{model:simu}, showing the periodic orbit in the plane
$\mathbb{R}^2$ using coordinates
\ba
\mathbf{y}=(y_1,y_2)=\left(x_a+\frac{1}{2}x_b,
\frac{\sqrt{3}}{2}x_b\right). \label{eqn:ycoords}
\ea
These, and other definitions for our coordinate systems, are illustrated in
figure~\ref{limit_cycle_example}.
Let $\Gamma \subset \mathbb{R}^2$ be the set of points on the limit cycle,
i.e. the red circle in figure~\ref{limit_cycle_example}. 
We further define the polar coordinates $(\rho,\theta)$ based on the centre
of the triangle, in the $\mathbf{y}$ coordinates, i.e. let
\ba
\rho(\mathbf{y}) = \sqrt{\left(y_1-\frac{1}{2}\right)^2
+ \left(y_2-\frac{\sqrt{3}}{6}\right)^2}, \nn
\ea
and
\ba
\theta(\mathbf{y}) = \arctan\left(\frac{y_1-1/2}{\sqrt{3}/6-y_2}\right). \nn
\ea
Let $T$ be the period of limit cycle, i.e. $T$ is the smallest
positive value such that $\forall \mathbf{y}\in \Gamma$, $\mathbf{y}(t)=\mathbf{y}(t+T)$ but for any $0<T^{'}<T$, $\mathbf{y}(t)\neq\mathbf{y}(t+T^{'})$.
All points in the interior of the limit cycle
(except the equilibrium point at $\bfy=(1/2,{\sqrt{3}}/{6})$ are
attracted to the limit cycle, enabling us to define a `landing point' on $\Gamma$
to which they are asymptotically attracted. Although the $\omega$-limit set
of a point $\bfy_0 \in \bbR^2$ would clearly be the entire orbit $\Gamma$,
by looking at the sequence defined
by advancing for multiples of the period $T$ we can identify a single
limit point $p_{\infty}(\mathbf{y}_0)$. Specifically we write the time-evolution
map for the ODEs as
$\phi_t(\mathbf{y}_0):=\mathbf{y}(t)$
where $\mathbf{y}(t)$ solves the ODEs~\eqref{eqn:rate_eq}
subject to the initial condition $\mathbf{y}(0)=\mathbf{y}_0$.
Then we define the sequence $\{p_n \}_{n \geq 0}$ by 
\ba
p_n=\phi_{nT}(\mathbf{y}_0),\quad \text{and} \quad p_0=\mathbf{y}_0. \nn
\ea
and the limit point
\ba
p_{\infty}(\mathbf{y}_0):=\lim_{n\to\infty} \phi_{nT}(\mathbf{y}_0). \nn
\ea
Note that $p_\infty(\bfy_0) \in \Gamma$ always, and that
if $\mathbf{y}_0\in \Gamma$ then $p_{\infty}(\mathbf{y}_0)=\mathbf{y}_0$.
    
\begin{figure}[!ht]
\bc
\includegraphics[width=0.7\textwidth]{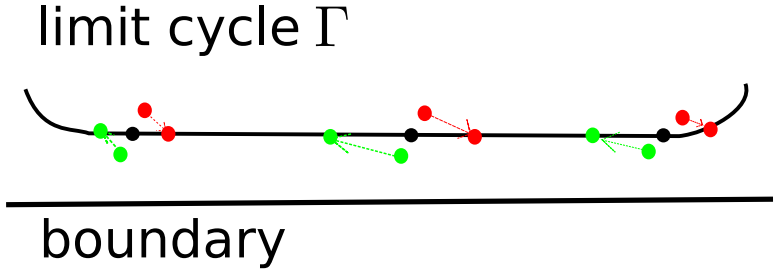}
\caption{Illustration of the idea of the asymptotic phase of small perturbations, using
the lower part of $\Gamma$ as shown in the $(y_1,y_2)$ plane.
Each of the three black points lies on the limit cycle $\Gamma$: trajectories
evolve along $\Gamma$ from left to right in the figure.
Perturbations of these points (in green) towards the boundary lead
to trajectories that have longer periods than $\Gamma$ has,
hence the green arrows indicate the asymptotic convergence of trajectories
back to the limit cycle to points that lie to the left of the
black dots. In contrast, perturbations (red dots) towards the interior
of $\Gamma$ lead to states that are accelerated by trajectories and
converge asymptotically to points on $\Gamma$ that are ahead of the black dots.}
\label{time_accelerate_delay}
\ec
\end{figure}

We can now define the asymptotic phase $\varphi(\mathbf{y}_0)$ of a point
near, but not necessarily on, the limit cycle by setting
\begin{equation}
\label{definition_phi}
\varphi(\mathbf{y}_0):=\theta(p_{\infty}(\mathbf{y}_0)).
\end{equation}
So that if $\bfy_0 \in \Gamma$, then $\varphi(\bfy_0)=\theta(\bfy_0)$, and curves
on which $\varphi(\bfy_0)$ is constant cross through $\Gamma$ at these points.
We can use $\varphi(\bfy_0)$ to consider the influence of noise, which pushes trajectories
off the limit cycle, causing time advances or delays, as illustrated in
figure~\ref{time_accelerate_delay}. This concept has been explored recently for a more general class of limit cycles in \cite{newby_effects_2014}, whose approach we follow here. It is analogous to the noise-induced drift observed in various ecological models including invasions \cite{Parsons2007b} and the evolution of altruism \cite{Constable2016}; see \cite{Parsons2015} for an introduction. 

The three-fold rotation symmetry of the problem suggests that it is
enough to focus on the interval $\varphi\in\left[-{\pi}/{3},{\pi}/{3}\right]$.

\subsection{A stochastic differential equation for $\varphi$}

Since $\varphi$ is a function on the phase space, we can derive an SDE for the
evolution of $\varphi$ using~\eqref{eqn:sde} and Ito's formula 
\cite{gardiner_stochastic_2009}. We obtain
\ba
\mathrm{d}\varphi(\mathbf{x}) & = & \left\{\sum_{i}\mathbf{A}_i(\mathbf{x},t)\partial_{i}\varphi(\mathbf{x}) + \frac{1}{2N}\sum_{i,j}\left[\mathbf{B}(\mathbf{x},t)\right]_{ij}\partial_{i}\partial_{j}\varphi(\mathbf{x})\right\}\mathrm{d}t \nn \\
& & + \frac{1}{\sqrt{N}} \sum_{i,j}\mathbf{G}_{ij}(\mathbf{x},t)
\partial_{i}\varphi(\mathbf{x})\mathrm{d}\mathbf{W}_{j}(t).
\label{eqn:phi_sde}
\ea
Note that our notation $\varphi(\bfx)$ really means $\varphi(\bfy(\bfx))$ since
$\varphi$ is defined by~\eqref{definition_phi} which uses the coordinates $\bfy$
defined in~\eqref{eqn:ycoords}. 
The advection vector $\mathbf{A}$ and the matrix
$\mathbf{B} \equiv \mathbf{G}\mathbf{G}^{T}$ are those given previously
in~\eqref{FPE:particle_drift} and~\eqref{FPE:particle_diffusion}.
 
\begin{figure}[!ht]
\bc
\includegraphics[height=110pt, trim=0 -14 75 0, clip=true]{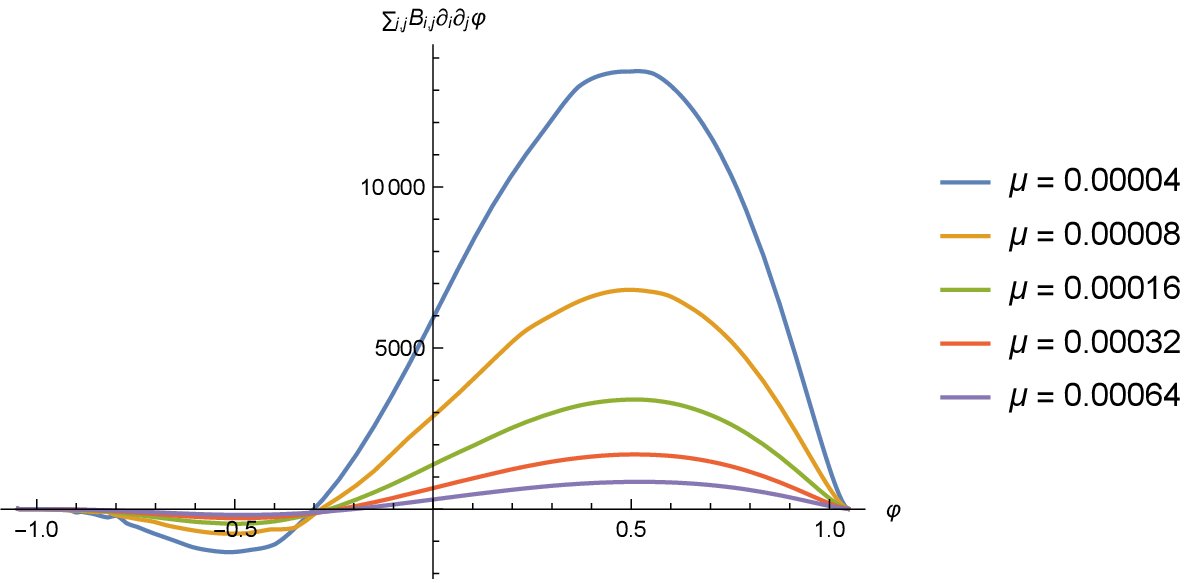}\includegraphics[height=110pt]{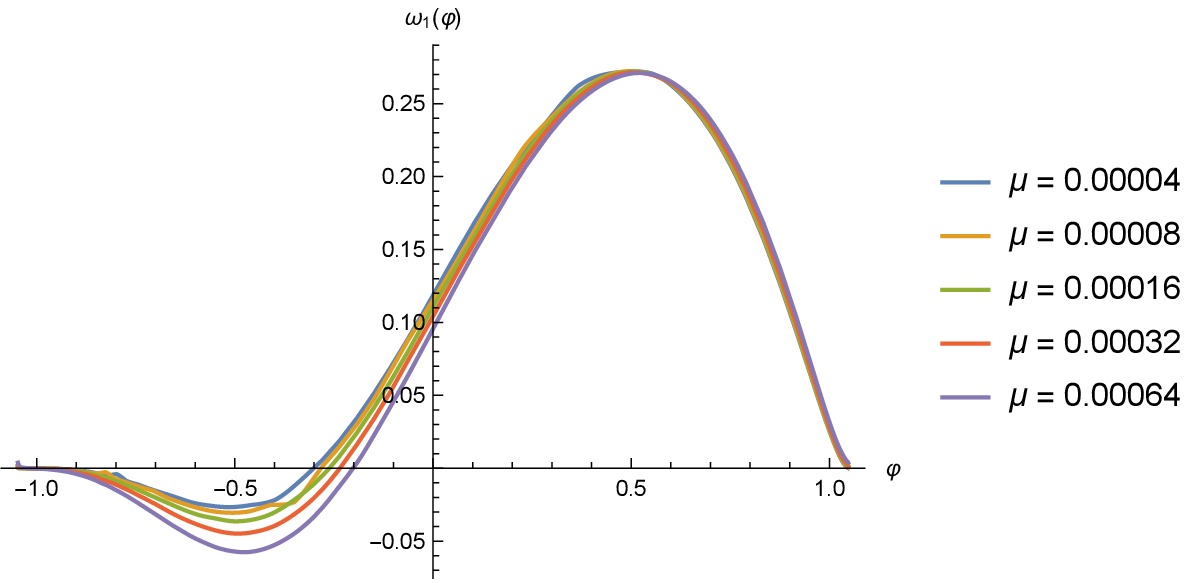}
\caption{Left: Numerically-computed functions $\sum_{i,j}\left[\mathbf{B}(\mathbf{x},t)\right]_{ij}\partial_{i}\partial_{j}\varphi$ for five different values of $\mu$. Right: Numerically-computed function $\omega_1(\varphi)$, the data collapse visible here justifies the inclusion of the explicit factor of $\mu$ in the defintion~{\eqref{eqn:omega1}}. Note that $\omega_1$ appears to cross through zero close to $\varphi=\pm \pi/3$.}
\label{fig:omega1}
\ec
\end{figure}

Although analytic expressions for $\varphi$ and its first and second
derivatives in phase space, i.e.
${\partial_a \varphi}$, ${\partial_b \varphi}$,
${\partial^2_a \varphi}$,
${\partial_a\partial_b \varphi}$,
and ${\partial^2_b \varphi}$ are unknown, they can be
estimated numerically.
Since we are interested in the behaviour of perturbations near $\Gamma$, we
have estimated these derivatives at points on $\Gamma$ and
then used them to define two functions of $\varphi$:
\ba
\omega_\infty (\varphi(\bfx)) & := & \sum_{i}\mathbf{A}_i(\mathbf{x},t)
\partial_{i}\varphi(\mathbf{x}), \quad \text{and} \nn \\
\omega_1(\varphi(\bfx)) & := & -\frac{\mu}{2}\sum_{i,j}\left[
\mathbf{B}(\mathbf{x},t)\right]_{ij}\partial_{i}\partial_{j}\varphi(\mathbf{x}).
\label{eqn:omega1}
\ea
Note that we have included in the definition of $\omega_1$ a prefector of $\mu$. This is necessary since the stochastic slowdown effect is stronger for smaller $\mu$, as indicated by the previous stochastic analysis, and shown numerically in the left panel of figure~{\ref{fig:omega1}}. We can conclude that, curiously, whilst the noise itself has negligible strength (order $1/N$) in region \II, the noise-induced slowdown has an order one impact on the dynamics. Specifically, the leading order form of equation~{\eqref{eqn:phi_sde}} in large $N$ and small $\mu$ is simply
\begin{equation}
\frac{\mathrm{d}}{\mathrm{d}t}\varphi = \omega_{\infty}(\varphi) 
- \frac{1}{N\mu}\omega_1(\varphi), \label{eqn:dphi}
\end{equation}
which enables us investigate the contributions to the period of the limit cycle
of the advection term $\bfA$ and the fluctuation-related contribution $\bfB$,
separately.

\begin{figure}[!ht]
\bc
\includegraphics[width=0.9\textwidth]{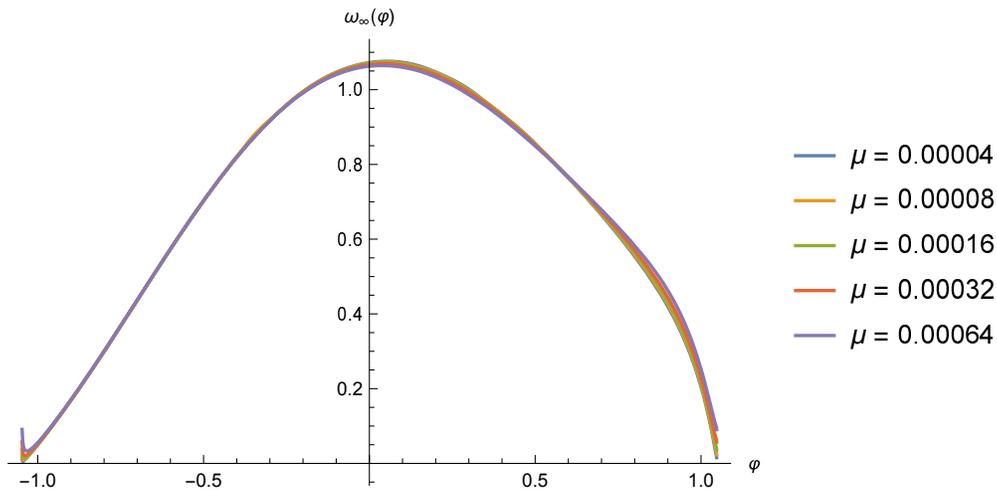} 
\caption{Numerically-computed function $\omega_{\infty}(\varphi)$ for five
different values of $\mu$ from $5\times 10^{-5}$ up to $8\times 10^{-4}$.}
\label{fig:omega_inf}
\ec
\end{figure}

Figures~\ref{fig:omega_inf} and~\ref{fig:omega1} show numerically-computed
approximations to the two functions $\omega_\infty(\varphi)$ and $\omega_1(\varphi)$,
respectively.
We observe that $\omega_\infty$ remains positive as $\varphi$ increases,
with a maximum near, but not exactly at, $\varphi=0$. This shows 
that points on $\Gamma$ move in the direction of increasing $\varphi$;
there is no reason for any symmetry about $\varphi=0$ due to the cyclic
nature of the dynamics. The shape of the curve $\omega_\infty (\varphi)$ 
varies little with $\mu$ except near the
equilibrium points in the corners, near $\varphi=\pm \pi/3$.
Figure~\ref{fig:omega1} shows the variation in $\omega_1(\varphi)$ as $\mu$
increases. Note that the definition of $\omega_1$ in ~\eqref{eqn:omega1} contains
a factor of $\mu$: including this factor of $\mu$ yields curves for 
$\omega_1(\varphi)$ that are extremely close to each other even as
$\mu$ increases by a factor of 16. Note also that the curves tend to
zero at both ends, indicating that the changes in $\mu$ do not change
the contribution from $\omega_1(\varphi)$ to the time spent near each
equilibrium point.
In the first part of the plot, $\omega_1(\varphi)$ is negative, indicating
that the angular velocity here overall, from~\eqref{eqn:dphi}, is increased, so
that the period of the orbit would be decreased. But this effect is more
than compensated for by the behaviour in the second part of the plot
where $\omega_1(\varphi)$ becomes more strongly positive, leading to a decrease in the overall
angular velocity here, and hence a larger increase in the period of the orbit.
The combined effect is therefore to increase the period of the orbit, and
this increase is driven directly by the fluctuations described by the
matrix $\bfB$ in the original SDE.

Equation~\eqref{eqn:phi_sde} can be used to quantify these
contributions to the period of the orbit since the
expected period of the limit cycle can be computed from the
SDE~\eqref{eqn:phi_sde} by writing
\begin{equation}
\label{T_SDE_1}
T_{\text{SDE}}=3\int_{-\frac{\pi}{3}}^{\frac{\pi}{3}}
\frac{\mathrm{d}\varphi}{\dot{\varphi}}
= 3\int_{-\frac{\pi}{3}}^{\frac{\pi}{3}}\frac{\mathrm{d}\varphi}{\omega_{\infty}(\varphi)-\frac{1}{N\mu}\omega_1(\varphi)},
\end{equation}
which, considering the case $N \gg 1$, can be approximated by
\ba
T_{\text{SDE}}=
3\int_{-\frac{\pi}{3}}^{\frac{\pi}{3}}\frac{\mathrm{d}\varphi}{\omega_{\infty}(\varphi)}
+\frac{3}{N\mu}\int_{-\frac{\pi}{3}}^{\frac{\pi}{3}}\frac{\omega_1(\varphi)}{\omega^2_{\infty}(\varphi)} \mathrm{d}\varphi + O\left( \frac{1}{N^2} \right). \nn
\ea
The first integral on the right hand side corresponds to the ODE approximation
$T_\text{ODE} \approx 3 \ln (1/\mu)$,
valid for larger $\mu$; we infer that $\omega_\infty(\varphi)$ depends
(relatively weakly) on $\mu$. The second integral contains the leading-order
$N$-dependent behaviour and shows that this is a contribution to the period
that scales as $T \propto 1/(N\mu)$. Although $\omega_\infty$ depends on $\mu$,
as observed above, this dependence is visible most obviously near to the
equilibrium points at which points $\omega_1(\varphi)$ is close to zero. Overall
we might therefore expect the $\mu$-dependence of the second integral to be
evn weaker than than $\ln \mu$ dependence of the $\omega_\infty(\varphi)$ function
itself. If this were the case, we would be left with just the $T \propto 1/(N\mu)$
dependence that would come to dominate the expression for $T_\text{SDE}$ for
small $\mu$.

In this way we observe that the SDE is able to capture the effects
of fluctuations near the limit cycle and yields an expression for the
period of oscillations that indicates that the period increases as $\mu$
decreases, leading into region \III-type scaling at small $\mu$. Moreover,
at larger $\mu$ (and larger $N$ at fixed $\mu$) we observe that the period
of oscillations is given by the deterministic expression computed
for region \I. Hence this analysis of region \II\ is able to capture the
cross-over in scalings for the periodic orbit as we move between regions \I\
and \III.

\section{Discussion}
\label{sec:discussion}

We have presented a detailed analysis of the simple Rock--Paper--Scissors
game, played in a well-mixed population, where the effects of cyclic competition
and mutation lead to oscillatory dynamics. In the mean-field ODE model where
the strategy mix in the population is described by replicator equations,
there is a stable limit cycle, produced in a Hopf bifurcation, when
the mutation rate $\mu$ is small enough compared to the asymmetry $\beta$
in the pay-off matrix. For large finite population sizes $N$, the dynamics
of the finite population, as given by stochastic `chemical reactions'
between the species, closely follow the ODE dynamics. Here we investigated 
a specific chemcial reaction scheme that was chosen to be straightforward to 
implement in the parameter regime we are interested in. It should be noted 
that this choice is not unique and may well not be optimal; for example, 
it does not allow for the possibility of $\beta<0$. 

Stochastic simulations in the regime of very small mutation rates
shows qualitatively, and quantitatively, different dynamics for the
oscillations, with the system remaining at corners for most of the time. Mutations
are rare, but an essential part of the dynamics. We presented a detailed
description of the dynamics in this regime using a Markov chain model
for the dynamics on the boundary of phase space, and then arguing that
in this regime the assumption that the dynamics took place on the boundary was
valid, and led to a self-consistent picture.

\begin{figure}[!ht]
\bc
\includegraphics[width=0.9\textwidth]{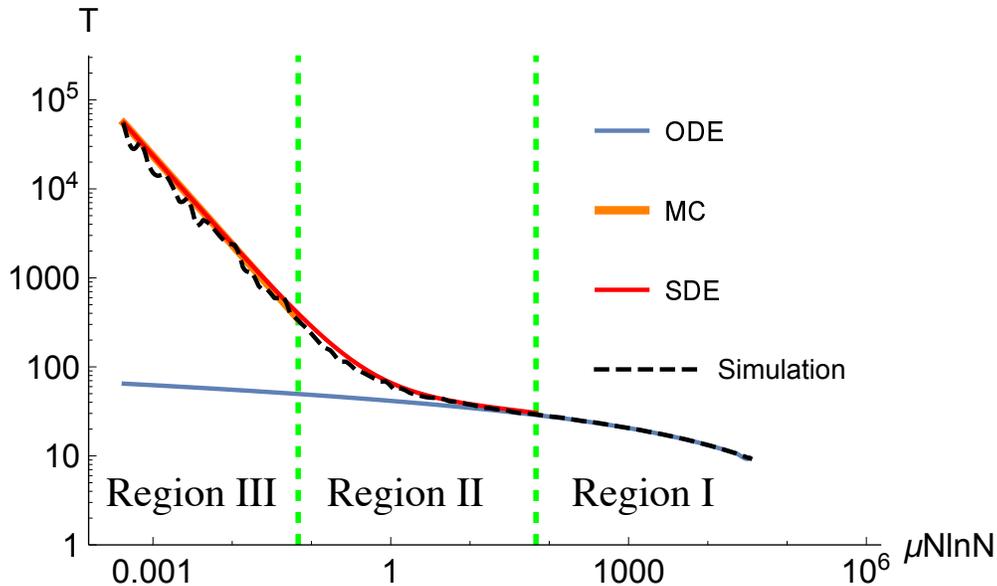}
\caption{A summary of the three regions for the dynamics, comparing the
theoretical predictions for the mean period $T$ of oscillations each region with the results of
stochastic simulations for $N=2^{17}$.}
\label{fig:summary}
\ec
\end{figure}

Our third approach to the problem was to construct a stochastic differential
equation, which, according to a theorem of Kurtz \cite{kurtz_strong_1978} approximates the
individual-based dynamics for large but finite populations. This was successful
in capturing the cross-over between the mean-field and fully stochastic
regimes as the mutation rate $\mu$ decreased for a fixed population size.
In general the effects of noise on nonlinear oscillators are complex
and topic of significant current research interest, see \cite{newby_effects_2014}.
Our work here has concentrated on understanding the cross-over between
regions \I\ and \III.
The cross-over occurs when $\mu N \ln N \approx 1$, and the period of the
oscillations changes from the mean-field
approximation $T_\text{ODE} \sim 3 \ln (1/\mu)$ to the stochastic
approximation $T_\text{SSA} \sim 3/(N\mu)$. Figure~\ref{fig:summary}
summarises these three regions of the behaviour, plotting the period
of the oscillations as a function of $\mu N \ln N$, comparing
the expected period of stochastic simulations with $N=2^{17}$ with
the theoretical results presented above.
 
Our main conclusion is that as demographic fluctuations become
more important (i.e. at small
mutation rates and for small population sizes) the period of the
cyclic oscillations increases significantly above that which would
be predicted on the basis of the mean-field ODE model.

To fully assess the importance of this effect in real populations will require further work. The presence of three-body interactions in our chemcial reaction scheme could be viewed as less biologically realistic, and some theorists would prefer a stochastic description in terms of a birth-death process with fitness and weak selection. Similarly, we considered here a model with fixed population size, yet, in the wild, population size is of course variable. One promising route to addressing these modelling concerns has recently been presented in {\cite{Constable2017}}, providing a robust mapping from Lotka-Volterra population models with two-body interactions and varying populations into their stochastic dynamics in frequency space. We would argue, however, that the salient features of the stochastic slowdown we observed should be common to all reasonable model specifications: in region I the dynamics are dominated by waiting for rare mutation events, and in regions II and III the format of the noise is subdominant to the deterministic features of the limit cycle. A more exciting direction for future work is to explore the role of the spatial distribution of populations, for example, one might ask if stochastic slowdown also affects spiral waves (see e.g. \cite{Szczesny2013,Szczesny2014}).

Our very detailed work here should also provide a basis for investigation of
many similar, more complex, kinds of cyclic interaction. We also intend
to examine situations where the mean-field behaviours are known to
produce more complicated dynamics, for example through
the occurrence of heteroclinic networks, or other bifurcations from
the heteroclinic cycles involved. This work then will provide a starting
point to guide the investigation of these situations when stochastic
effects due to large but finite populations of individuals become important.

\section*{Acknowledgement}
QY is supported by the University of Bath through a 50th Anniversary
Excellence Award. TR and JHPD acknowledge support from the Royal Society.

\end{document}